\documentstyle[12pt]{article}
\begin{document}
\begin{center}
{\Large\bf Moment Analysis of Multiplicity Distributions}

\bigskip

{\large A.~Capella, I.M.~Dremin\footnote{Permanent address: P.N.~Lebedev
Physical Institute, Moscow, Russia},
V.A.~Nechitailo$^1$, J.~Tran Thanh Van}

\bigskip

{\it Laboratoire de Physique Theorique et Hautes Energies,\\
Bat. 211, Universite Paris-Sud, 91405 Orsay, France}
\end{center}

\begin{abstract}
Moment analysis of global multiplicity distribution previously done for
$e^{+}e^{-}$ and $hh$ processes is applied to $hA$ and $AA$ collisions.
The oscillations of cumulants as functions of their rank are found in
all the cases. Some phenomenological approaches and quark-gluon string
models are confronted to experimental data. It has been shown that the
analysis is a powerful tool for revealing the tiny features of the
distributions, and its qualitative results in various processes are
rather stable for different multiplicity cut-offs determined by
experimental (or Monte-Carlo) statistics.
\end{abstract}

\section{Introduction}
Multiplicity distributions of multiparticle reactions contain in the
integrated form all the correlations of the system. At the same time
they are measured with best accuracy in experiment. Therefore their
study is of particular interest. The shapes of the distributions differ
drastically for various processes at different energies.
Phenomenologically, it is popular to use Poisson (or even sub-Poisson)
distributions at lower energies and negative binomial distribution
(nowadays, modified negative binomial distribution) at higher energies.
The parameters of the distributions vary for different cases and are not
determined with high enough precision when fits are done.

However, any multiplicity distribution can be represented not only by
the probabilities of $n$-particle events but also by its moments or by
its generating function. It happens that such description is preferred
both from the theoretical point of view and for the stability of results
obtained from experimental data for various reactions. The solution of
QCD equations for generating functions predicts \cite{d1,d2} the
oscillatory behaviour of the cumulant moments of the {\it parton}
distribution in QCD jets as functions of their rank with first
minimum located at the point determined by the inverse power of the
anomalous dimension $\gamma $.

Surprisingly enough, the oscillations of the ratio of cumulant to
factorial moments with similar periodicity have been found in
experimental {\it hadron} multiplicity distributions in $e^{+}e^{-}$
and $pp(p\bar p)$ collisions at high energies \cite{d3}. No pattern
of this kind appears in phenomenological fits (for review see
\cite{d4}).

In this paper we apply the moment analysis to multiplicity distributions
in hadron-nucleus and nucleus-nucleus collisions. The similar (to $ee$
and $hh$) shape of the above ratio appears in both cases. The only
difference is in the depth of the first minimum for different
projectiles and targets but not in its location. We compare these
results to purely phenomenological fits of the distributions and to some
theoretical models (DPM - dual parton model and QGSM - quark-gluon
string model). While phenomenological fits do not describe shapes of the
ratio obtained from experimental data, the string models reproduce them
qualitatively rather well. The minor quantitative difference can be
accounted for by slight variation of parameters and should be attributed
to experimental uncertainties that demonstrate also how sensitive the
method of moment analysis is to tiny features of multiplicity
distributions. In that aspect, the qualitative stability of the ratio
shapes for reactions with drastically different distributions looks very
impressive. The general feature of branching processes inherent both in
QCD and in theoretical Monte-Carlo models can be in charge of it.

We have checked that the truncation of multiplicity distributions due to
finite statistics and conservation laws does not influence drastically
this dependence even though it produces some additional oscillation
effect which should be taken into account when comparing models to
experimental data. The theoretical arguments in favor of it can be got
from studies in paper \cite{mp5}, where it has been shown that
conservation laws give rise to corrections of the order of $\gamma ^3$,
while the oscillation effect appears already due to modified leading
logarithm approximation terms \cite{d1} of the order of $\gamma ^2$.

Besides, we have found the zeros of the truncated generating functions.
Similar to $ee$ and $hh$ processes, they tend to lie close to the circle
of unit radius in the complex plane for the high enough truncation
limit. Since the highest multiplicities are in $AA$-collisions, the
circle is most clearly seen there. The rightmost zeros tend to the real
axis at higher multiplicities that reminds of Lee-Yang conjecture in
statistical physics about the phase transition point to which the
rightmost zeros of the grand partition function approach when volume
increased. It implies that the singularity of the generating function
is very close to the point where all the moments are calculated as
derivatives of the generating function. That explains why the moment
analysis is so sensitive to slight variations of multiplicity
distributions and, at the same time, provides qualitatively similar
results for various processes.

\section{Moment Analysis of Hadron-Nucleus and Nucleus-Nucleus
Collisions}

First, let us describe the generalities of the moment analysis. The
normalized factorial $(F_q)$ and cumulant $(K_q)$ moments of the
multiplicity distribution $P_n$ are defined as
\begin{equation}
F_q = \frac {\langle n(n-1)\ldots (n-q+1)\rangle }{\langle n\rangle ^q}
\equiv \frac {\sum _{n=0}^{\infty }n(n-1)\ldots (n-q+1)P_n}
{(\sum _{n=0}^{\infty }nP_n )^q} ,  \label{1}
\end{equation}
\begin{equation}
F_q = \sum _{m=0}^{q-1} C_{q-1}^{m}K_{q-m}F_m , \label{2}
\end{equation}
where $P_n$ is a probability of $n$-particle events, $q$ is the rank of
the moment, $C_{n}^{m}=n!/m!(n-m)!$ are the binomial coefficients.

Thus, if the multiplicity distribution $P_n$ is known one calculates the
factorial moments of any rank according to (\ref{1}), and, afterwards,
using the recurrence relations (\ref{2}) finds out the cumulants.

Theoretically, it is more convenient to start with the generating
function
\begin{equation}
G(z) = \sum _{n=0}^{\infty }z^{n}P_n , \label{3}
\end{equation}
and calculate the moments as its derivatives
\begin{equation}
F_q = \left. \frac {1}{\langle n\rangle ^q}\frac {d^{q}G}{dz^q}\right| _{z=1} ,
\label{4}
\end{equation}
\begin{equation}
K_q = \left. \frac {1}{\langle n\rangle ^q}\frac {d^{q}\ln G}{dz^q}\right|
_{z=1}. \label{5}
\end{equation}
In practice, the multiplicity distribution is known up to some maximum
multiplicity $N$, and one has to deal with the truncated generating
function
\begin{equation}
G_{N} = \sum _{n=0}^{N}z^{n}P_n. \label{6}
\end{equation}
The interest to the moment analysis of multiplicity data appeared after
the solution of QCD equations for the generating function was found
\cite {d1,d2}. It predicts a special oscillation pattern for the ratio
of cumulant to factorial moments
\begin{equation}
H_q = \frac {K_q}{F_q} \label{7}
\end{equation}
with first minimum located at $q_{min}\approx 5$ and determined by the
inverse value of the QCD anomalous dimension
\begin{equation}
\gamma _{0} =(2N_{c}\alpha _{s}/\pi )^{1/2} , \label{8}
\end{equation}
where $\alpha _s$ is a coupling constant, $N_{c}=3$ is the number of
colours. For further details see papers \cite{d1,d2} and the review
paper \cite{d4}.

Surely, it was the prediction for the moments of {\it parton} (mostly,
gluon) multiplicity distributions. However, when applied to final
{\it hadrons} in $e^{+}e^{-}$ and $pp(p\bar p)$ experimental events
\cite{d3} the analysis shows the similar structure of the ratio $H_q$.

We extend this analysis to hadron-nucleus and nucleus-nucleus collisions
using experimental data, theoretical Monte-Carlo models and
phenomenological fits. We check also how important are various
multiplicity cut-offs for the moments. The zeros of the truncated
generating function (\ref{6}) at various cut-offs $N$ are found and
discussed.

This analysis is of a special interest because the shapes of
multiplicity distributions themselves differ strongly from those in
$ee$ and $pp$ collisions.

In Fig.1 we plot the ratio $H_q$ calculated according to the dual parton
model \cite{6} and quark-gluon string model \cite{7} for hadron-nucleus
and nucleus-nucleus collisions. For a comparison, we show this ratio
for charged particles in proton-antiproton collisions at 546 GeV
calculated from experimental data and published in \cite{d3}.

One concludes that the general shapes of $H_q$ are similar for all
these processes. The oscillations with first negative minimum in the
range $q=4-6$ are clearly seen. Its depth increases at each step from
$hh$ to $hA$ and $AA$. The stability of qualitative features of moments
is especially remarkable if one compares the multiplicity distributions
in all the cases which are very different, indeed.

No phenomenological fit, we are aware of, is able to reproduce such
a behaviour. For Poisson distribution all cumulants (and, consequently,
ratios $H_q$) are identically equal to zero (at $q\geq 2$). The modified
 negative binomial distribution has the generating function
\begin{equation}
G^{(MNBD)}(z) = \left( \frac {1+\Delta(1-z)}{1+r(1-z)}\right)^k
\label{9}
\end{equation}
with three free parameters $r,\Delta ,k$ related to the average
multiplicity by $\langle n\rangle = k(r-\Delta )$. The negative binomial
distribution is obtained for $\Delta =0$. Using (\ref{5}) it is easy to
show that
\begin{equation}
K_{q}^{(MNBD)} = k^{1-q}(q-1)!(r^{q}-\Delta ^{q})/(r-\Delta )^q .
\label{10}
\end{equation}
Since $k>0$, the cumulants are always positive for the negative binomial
distribution. Recent fits of $e^{+}e^{-}$ data \cite{8,9} give rise to
$\Delta <0$ and $r\leq |\Delta |<1$. According to (\ref{10}) it would
imply that cumulants change sign at each $q$ being negative at even
values of $q$ and positive at odd ones. They do not reveal the broad
oscillations shown in Fig.1. The whole picture reminds somewhat the
pattern of fixed multiplicity distribution (see \cite{d4}). In general,
we would like to comment that the fits used in \cite{9} look rather
strange since they imply sub-Poissonian distribution of negatively
charged particles (with $K_{2}^{(-)}<0$), while it is usually claimed
that it is super-Poissonian one at high energies. Moreover, with the set
 of parameters $\Delta =-0.75, r=0.65$ available in \cite {8,9} one
gets negative probabilities for large $n$ as is easily seen from
(\ref{9}) and was confirmed by computer calculations. Moment analysis
of charged particle distributions in $e^{+}e^{-}$ data done in \cite
{d3,10} has shown the pattern reminding that of $p\bar p$ and,
according to Fig.1, of reactions with nuclei. The only qualitative
difference is the smaller depth of the minimum in $ee$ compared to
hadronic reactions.

Contrary to phenomenological fits, the dual parton model \cite{6} and
the quark-gluon string model \cite{7} look quite successful in
describing the qualitative behaviour of the ratio $H_q$ in
hadron-nucleus collisions as shown in Figs.2-4, where the experimental
data of NA22 collaboration at 250 GeV have been used \cite{11}.
The general periodicity of the curves and the absolute values at
minima and maxima are reproduced rather well even though their
positions are somewhat shifted for pAl and KAl. There are several
factors which influence the shapes of $H_q$-curves, and we discuss them
in the next section.

\section{Asymptotically Subleading Terms of the Moments}

The stability of the oscillating pattern in Fig.1 is astonishing
because, first, the pattern differs from those of widely used in
probability theory and, second, the multiplicity distributions in
various processes under consideration are very different and do not seem
to have much in common (contrary to $H_q$-ratios). Moreover, this
pattern seems sometimes very sensitive to tiny modifications of
multiplicities themselves. This is related to the subtraction procedure
used to calculate cumulants according to eq. (\ref{2}) when factorial
moments are known.

However, when comparing to experimental data one should be sure that the
proper distributions have been chosen. By that, for example, we mean
that the moment analysis of charged multiplicities provides different
values of ratios $H_{q}^{(ch)}$ compared to the values $H_{q}^{(-)}$
obtained for negatively charged particles. If one considers $e^{+}e^{-}$
collisions, the number of charged particles is twice that of negatives.
Therefore, the corresponding generating functions are related by
\begin{equation}
G_{ch}(z) = G_{-}(z^2).  \label{11}
\end{equation}
Herefrom, one gets for the first five moments:
\begin{eqnarray}
F_{1}^{(ch)}=F_{1}^{(-)}=K_{1}^{(ch)}=K_{1}^{(-)}=1,
\nonumber \\
F_{2}^{(ch)}=F_{2}^{(-)}+\frac {1}{\langle n_{ch}\rangle },
\nonumber \\
F_{3}^{(ch)}=F_{3}^{(-)}+\frac {3}{\langle n_{ch}\rangle }F_{2}^{(-)},
\label{12} \\
F_{4}^{(ch)}=F_{4}^{(-)}+\frac {6}{\langle n_{ch}\rangle}F_{3}^{(-)}+
\frac {3}{\langle n_{ch}\rangle ^{2}}F_{2}^{(-)},
\nonumber \\
F_{5}^{(ch)}=F_{5}^{(-)}+\frac {10}{\langle n_{ch}\rangle}F_{4}^{(-)}+
\frac {15}{\langle n_{ch}\rangle ^{2}}F_{3}^{(-)},
\nonumber
\end{eqnarray}
and the same relations are valid for cumulants (one should just replace
$F$ by $K$ in (\ref{12})). One concludes that $F_{q}^{(ch)}>F_{q}^{(-)}$
at any $q$, while the inequality $K_{2}^{(ch)}>K_{2}^{(-)}$ is always
valid for cumulants but $K_{3}^{(ch)}$ can be less than $K_{3}^{(-)}$
if the distribution of negatives is sub-Poiisonian one (i.e., if
$K_{2}^{(-)}<0$). If negative particles are distributed according to
Poisson law i.e. $K_{q}^{(-)}=0$ for $q\geq 2$, the charged particles
dispersion differs from Poissonian. In that case, $K_{2}^{(ch)} =
\langle n \rangle ^{-1} >0$ but higher cumulants are equal to zero
according to (\ref{12}) as if Poisson law is restored just for them.

One can think about this procedure as if the negative particles are
created by a ``cluster'' of negative+positive particle (due to charge
conservation they are always produced in pairs). In general, one can
consider a cluster model with a definite probability for cluster
production and its decay into $k$ particles (see \cite{12}, for
example). In that case the relation between the generating functions for
particles  and clusters is
\begin{equation}
G_{p}(z)=G_{c}(z^k ),  \label{13}
\end{equation}
and the formulae (\ref{12}) can be easily generalized to
\begin{eqnarray}
F_{2}^{(p)}=F_{2}^{(c)}+\frac {k-1}{\langle n_{p}\rangle}, \nonumber \\
F_{3}^{(p)}=F_{3}^{(c)}+\frac {3(k-1)}{\langle n_{p}\rangle}F_{2}^{(c)}+
\frac {(k-1)(k-2)}{\langle n_{p}^{2}\rangle} \label{14}
\end{eqnarray}
etc. Herefrom, one gets particle distributions wider than cluster ones
because $F_{2}^{(p)}>F_{2}^{(c)}$ for $k>1$, e.g., super-Poissonian ones
for particles created by Poisson clusters typical for asymptotical
multiperipheral and string models. In particular, for the model of
Poissonian clusters with $k=1.4$ charged particles per cluster \cite{12}
one gets
\begin{equation}
K_{2}^{(p)}=\frac {0.4}{\langle n_{p}\rangle}>0; \,\,
K_{3}^{(p)}=-\frac {0.24}{\langle n_{p}\rangle ^2}<0. \label{15}
\end{equation}
To get $K_{3}^{(p)}>0$, one needs $k>2$.

If KNO-scaling holds at asymptotically high energies, all $F_q$ tend to
constants and $\langle n_{ch}\rangle \rightarrow \infty $. Therefore,
all the moments of both distributions coincide at asymptotics. At
present energies, however, the correction terms of the order
$O(1/\langle n_{ch}\rangle )$ can be still important, especially, for
those cumulants which are close to zero. Therefore, cumulant analysis
can give different results for different distributions. Above, we
applied it to charged particles multiplicities.

The finite energy of colliding particles and the final experimental
statistics truncate the multiplicity distribution at some finite
multiplicity $N$. The relations (\ref{12}), (\ref{14}) are independent
of this truncation since they are valid for truncated generating
functions (\ref{6}) as well as for total ones (\ref{3}). However, it
influences the values of the moments in these formulae. E.g., the
truncated Poisson law for negative particles would give rise to non-zero
values of cumulants. Actually, it was conjectured in \cite{13,14} that
at present energies this effect could be rather important for $ee$ and
$pp$ collisions since it imitates the oscillations of cumulants. Surely,
it should disappear at higher energies. This is clear both from model
calculations and from theoretical arguments of \cite{mp5} where this
effect is of the order of $\gamma ^3$, while the oscillations in QCD
appear already at the modified logarithm approximation of $\gamma ^2$
order \cite{d1,d4}. Nevertheless, at present energies it should be
carefully treated for each experiment together with uncertainties
imposed by the error bars due to finite statistics. In any Monte-Carlo
model they can be reduced by enlarged statistics which is a matter of
computer time only.

We do not show the error bars in our Figs but the stability of curves
and the regularity of trends shows that they hardly can change our
conclusions. For $ee$ processes the minima are smaller but even there it
was shown \cite{d3,10} that the first minimum is still reliable.
We show in Fig.5 that the shapes of $H_q$-curves in nucleus-nucleus
collisions are not very sensitive to the multiplicity cut-off.
Truncating the rather flat multiplicity distribution at two different
values of $N$ we do not observe any strong shift of the curve and
noticeable change of its shape.

\section{Zeros of the Truncated Generating Function}

The truncated generating function (\ref{6}) is a polynomial of $N$-th
power with positive coefficients. Therefore, it possesses $N$ complex
conjugate zeros in the complex $z$-plane. We have found their locations
in all the above cases varying the value of $N$.

The general picture becomes quite stable at large $N$. The zeros tend to
lie close to unit circle in $z$-plane. The rightmost ones move to the
real axis with $N$ increasing as if trying to pinch it at some value of
$z$ slightly exceeding 1. They move closer and closer to 1 if we compare
$ee, hh, hA, AA$ collisions, correspondingly. Their convergence point
implies that the singularity of the total generating function is located
near $z=1$, where all the moments are calculated (see eqs.
(\ref{4}),(\ref{5})). It means that moments are very sensitive to the
location and the origin of the singularity. (Let us note that the
singularity of the negative binomial distribution appears at
$z=1+r^{-1}$ according to (\ref{9}), i.e. quite far from $z=1$ if $r<1$
as proposed in \cite{8} that contradicts to Figs below).

We demonstrate all these features in Figs.6-9. To show how close are the
rightmost zeros to the point $z=1$ in different processes we give their
coordinates $x, y$ and the radii $R=(x^2 +y^2 )^{1/2}$ in the Table.
Also shown are the maximum multiplicities $N$. All the radii exceed
slightly 1, and the nearest ones to 1 are in nucleus-nucleus collisions.

The above findings could be interesting also if one speculates about the
statistical analogies in particle physics \cite{15} in terms of
Feynman-Wilson liquid \cite{16,17}. Then the truncated generating
function reminds the grand partition function and the variable $z$ plays
the role of fugacity. It has been proven by Lee and Yang that its zeros
should lie on the circle in the complex plane and the tendency of
rightmost zeros to pinch the positive real axis means that there is a
phase transition in the system. Let us note, however, that there is no
symmetry in $P_n$ which helped Lee and Yang to prove their statement
about zeros location. The similar patterns appearing in all processes in
particle physics are quite encouraging for such analogies. The nature of
the singularity could be revealed from analysis of experimental data, in
principle.

\section{Discussion and Conclusions}

The moment analysis of multiplicity distributions as applied to $hA$ and
$AA$ collisions reveals the oscillation pattern of the ratio $H_q$
similar to that previously found for $ee$ and $hh$ processes. However,
the amplitude of oscillations increases for heavier colliding objects,
while their periodicity remains stable. Phenomenological fits do not
show anything similar to observed patterns. The string models reproduce
them rather well. The multiplicity cut-off is not very important until
one comes too close to average multiplicity. The zeros of the truncated
generating function tend to be near the unit circle in the complex
plane with rightmost zeros moving at high multiplicities toward the real
positive axis and pinching it near the point $z=1$ where
derivatives are taken when moments are calculated. This is the
singularity point of the total generating function and it is closer to
$z=1$ for heavier colliding objects.

These findings show that there is much more similarity in different
collision processes than it could be envisaged from rather different
multiplicity distributions. Probably, it is due to the common branching
origin of them.

Also, the singularity point of the total generating function would be an
interesting topic to learn more about. The density of zeros and their
movement to the real axis at higher multiplicities in various processes
could indicate its nature.

Since the moments are calculated as derivatives of the generating
function at the point very close to the singularity, the moment analysis
is powerful and sensitive tool of revealing its structure and tiny
features of multiplicity distributions.

\section{Acknowledgements}
We are indebted to F.Rizatdinova for providing the experimental and
QGSM multiplicity distributions in tables.

This work was supported by INTAS grant 93-79 and by the Russian Fund for
Fundamental Research grant 93-02-3815.

\newpage

\noindent {\Large\bf Figure Captions}
\begin{description}
\item[Fig.1] The ratio $H_q$ for p$\bar{\rm p}$, hA and AA collisions.
             (See text for more details).
\item[Fig.2]  The ratio $H_q$ for pAl collisions at 250 GeV
              as calculated from experimental data \cite{11}
              and according to DPM \cite{6} and QGSM \cite{7}.
\item[Fig.3] The ratio $H_q$ for $\pi$Al collisions at 250 GeV
              (experimental data \cite{11} and QGSM \cite{7}).
\item[Fig.4] The ratio $H_q$ for $K$Al collisions at 250 GeV
              (experimental data \cite{11} and QGSM \cite{7}).
\item[Fig.5] The ratio $H_q$ for $^{32}$S$^{238}$U collisions
             at 200A GeV of $^{32}$S in the laboratory system,
             calculated from DPM multiplicity distributions
             truncated at two values of maximum multiplicity N.
\item[Fig.6] Zeros of the truncated generating function for pAl
             collisions at 250 GeV.
\item[Fig.7] The same as Fig.6 but for $\pi$Al.
\item[Fig.8] The same as Fig.6 but for $K$Al.
\item[Fig.9] Zeros of the truncated generating function for
             $^{32}$S$^{238}$U collisions
             at 200A GeV of $^{32}$S in the laboratory system
             according to DPM \cite{7}
\end{description}

\noindent {\Large\bf Table}\\
Location of the rightmost zeros of the truncated generating functions
for different processes.

\medskip

\begin{tabular}{|l|c|c|c|c|}
\hline
collision    & R        & y        & x      & N  \\
\hline
$\pi$ Al (QGSM)\ \ &\ \ 1.16042 \ \  &\ \ 0.24190 \ \
&\ \ 1.1349 \ \ &\ \ 41 \ \ \\
\hline
$\pi$ Al (exp)  & 1.16042  & 0.24771  & 1.1337 & 41 \\
\hline
$K$ Al (QGSM)  & 1.12553  & 0.28954  & 1.0877 & 32 \\
\hline
$K$ Al (exp)   & 1.14096  & 0.30445  & 1.0996 & 32 \\
\hline
p Al (DPM)   & 1.15537  & 0.29461  & 1.1172 & 27 \\
\hline
p Al (QGSM)  & 1.09168  & 0.28592  & 1.0536 & 27 \\
\hline
p Al (exp)   & 1.03598  & 0.31132  & 0.9881 & 27 \\
\hline
S U  (DPM)   & 1.02265  & 0.07061  & 1.0202 & 80 \\
\hline
\end{tabular}
\end{document}